\documentclass[11pt,preprint]{aastex}
\pdfoutput = 1
\newcommand{\Msun}{\mbox{$M_{\sun}$}}

\shorttitle{CO(40-39) and the AGN torus in NGC1068}
\shortauthors{Janssen et al.}

\usepackage{graphicx}
\usepackage{amsfonts}
\usepackage{amsmath}
\usepackage{natbib}
\usepackage{upgreek}
\usepackage{graphicx}
\usepackage{placeins}
\usepackage{url}
\usepackage{appendix}

\begin{document}

   \title{A deep Herschel/PACS observation of CO(40-39) in NGC 1068: \\ a search for the molecular torus
   }
\author{
A.W. Janssen\altaffilmark{1},
S. Bruderer\altaffilmark{1},
E. Sturm\altaffilmark{1},
A. Contursi\altaffilmark{1},
R. Davies\altaffilmark{1},
S. Hailey-Dunsheath\altaffilmark{2},
A. Poglitsch\altaffilmark{1},
R. Genzel\altaffilmark{1},
J. Graci{\'a}-Carpio\altaffilmark{1},
D. Lutz\altaffilmark{1},
L. Tacconi\altaffilmark{1}
J. Fischer\altaffilmark{3},
E. Gonz{\'a}lez-Alfonso\altaffilmark{4},
A. Sternberg\altaffilmark{5},
S. Veilleux\altaffilmark{6,7}
A. Verma\altaffilmark{8}
L. Burtscher\altaffilmark{1}
}

\altaffiltext{1}{Max-Planck-Institute for Extraterrestrial Physics (MPE), Giessenbachstra{\ss}e 1, 85748 Garching, Germany}
\altaffiltext{2}{California Institute of Technology, 301-17, 1200 E. California Blvd., Pasadena, CA 91125, USA}
\altaffiltext{3}{Naval Research Laboratory, Remote Sensing Division, 4555 Overlook Ave SW, Washington, DC 20375, USA}
\altaffiltext{4}{Universidad de Alcal{\'a} de Henares, 28871 Alcal{\'a} de Henares, Madrid, Spain}
\altaffiltext{5}{Tel Aviv University, Sackler School of Physics \&\ Astronomy, Ramat Aviv 69978, Israel}
\altaffiltext{6}{Department of Astronomy, University of Maryland, College Park, MD 20742, USA}
\altaffiltext{7}{Joint Space-Science Institute, University of Maryland, College Park, MD 20742, USA}
\altaffiltext{8}{Oxford University, Dept. of Astrophysics, Oxford OX1 3RH, UK}

\email{janssen@mpe.mpg.de}


\begin{abstract}
 {Emission from high-J CO lines in galaxies has long been proposed as a tracer
of X-ray dominated regions (XDRs) produced by AGN. Of particular interest is the question of whether the obscuring torus, which is required by AGN unification models, can be observed via high-J CO cooling lines. 

Here we report on the analysis of a deep Herschel-PACS observation of an extremely high J CO transition (40-39) in the Seyfert 2 galaxy NGC\,1068. The line was not detected, with a derived 3$\sigma$ upper limit of $2 \times 10^{-17}\,\text{W}\,\text{m}^{-2}$. We apply an XDR model in order to investigate whether the upper limit constrains the properties of a molecular torus in NGC 1068. The XDR model predicts the CO Spectral Line Energy Distributions for various gas densities and illuminating X-ray fluxes. In our model, the CO(40-39) upper limit is matched by gas with densities $\sim 10^{6}-10^{7}\,\text{cm}^{-3}$, located at  $1.6-5\,\text{pc}$ from the  AGN, with column densities of at least $10^{25}\,\text{cm}^{-2}$. At such high column densities, however, dust absorbs most of the CO(40-39) line emission at $\lambda = 65.69\, \upmu$m. Therefore, even if NGC 1068 has a molecular torus which radiates in the CO(40-39) line, the dust can attenuate the line emission to below the PACS detection limit. 
The upper limit is thus consistent with the existence of a molecular torus in NGC 1068. In general, we expect that the CO(40-39) is observable in only a few AGN nuclei (if at all), because of the required high gas column density, and absorption by dust.

}
\end{abstract}


   \maketitle
%

\section{Introduction}
Low-J rotational lines (up to J $\approx$ 7) of CO have long been used as a tracer for molecular gas.
With the Herschel Space Telescope\footnote{Herschel is an ESA space observatory with science instruments provided by European-led Principal Investigator consortia and with important participation from NASA.} \citep{Pilbratt-2010} it was possible to observe CO lines of higher J levels,  extending the CO Spectral Line Energy Distribution (SLED) to $\text{J} = 50$ (for $\text{J} \rightarrow \text{J} - 1$). The PACS instrument \citep{PACS-2010} on board Herschel samples the highest frequencies, with $13 < $ J $ < 50$, which we will call the high-J CO lines.  These lines with wavelengths $ 186.0\, \upmu\text{m}< \lambda < 52.85\, \upmu\text{m}$ arise from states $\sim 500-7,000\,\text{K}$ above ground, with critical densities of $\sim 10^{6}-10^{8}\, \text{cm}^{-3}$, and are thus good tracers of the Photon Dominated Regions (PDR) in the warm and dense Interstellar Matter (ISM) around stars, in shocks, or in the X-ray Dominated Regions (XDRs) of AGN.

AGN unification models require a structure that obscures the UV and optical emission from the Broad Line Region of Type 2 AGN from our view \citep{Antonucci-1985}. Further, compton thick AGN, like NGC 1068 \citep{Elvis-1988}, require an X-ray absorbing medium with a column density of at least $ N_{H} = 10^{24}\,\text{cm}^{-2}$, potentially in the  form of molecular gas. These two effects are often assumed to be caused by a dusty torus, although the gas and the dust are not necessarily distributed in the same way.  In this paper, we investigate whether high-J  CO lines can trace the X-ray absorbing gas (from now on called 'the torus'),  be it in the form of a classical torus \citep{KrolikBegelman-1988, KrolikLepp-1989} or of a more clumpy and more extended cloud structure. The CO emission from a parsec-scale torus was first modeled by \citet{KrolikLepp-1989} (KL89 hereafter), who found that the line fluxes should scale as J$^{3}$ up to J$\approx$ 58 (in this particular model for NGC 1068), and with absolute 
fluxes 
proportional to the total absorbed hard X-ray luminosity. More recent torus models have envisioned a dynamical structure of small clouds distributed over several parsecs surrounding the AGN \citep{Nenkova-2002, Hoenig-2006}. However, these newer clumpy torus models do not yet incorporate a gas phase and have been used so far only to predict the dust emission from an AGN.

In \citet{Hailey-Dunsheath-2012} (hereafter HD12) we carried out a detailed analysis of the full CO SLED of NGC 1068 up to ${\text{J}} = 30$ (for higher J only upper limits were found).
We concluded that two components are responsible for the CO SLED between $10 < J < 31$: 
the $\sim$200 pc diameter ring and the northern streamer, both traced by H$_{2}$~1-0~S(1) observations \citep{Mueller-2009}. Would the molecular torus, if it exists, be observable in the $J > 30$ lines? In order to answer this question, we carried out an additional, deeper, observation of CO(40-39).

The paper is organized as follows: a description of the existing multiwavelength observations of NGC\,1068 is given in section \ref{sec:1068}. Section \ref{sec:observations} covers the PACS observations and data reduction and presents the full (55 to 200 $\upmu$m) spectrum of the central spatial pixel (spaxel). Section \ref{sec:xdr-fit} describes updates on the PDR-, XDR-, and shock-fits presented in HD12, including the new upper limit on CO(40-39) and a small correction to the CO SLED. A brief description of the XDR code used to model the torus emission is given in section  \ref{sec:model}. A discussion of the results follows in section \ref{sec:discussion}.

\section{NGC\,1068}
\label{sec:1068}
NGC\,1068 is a Seyfert 2 galaxy at a distance of 14.4 Mpc, implying a 72 pc/arc second scale \citep{Bland-Hawthorn-1997}. Because of its proximity and luminosity, NGC\,1068 is a favorable object for testing hypotheses on Seyfert galaxies.

Early X-ray observations showed emission from the AGN nucleus, but only in scattered light \citep{Elvis-1988} because the AGN is completely obscured from our view by a column of $N_{H} = 10^{25}\,\text{cm}^{-2}$ \citep{Bauer-2014}. The intrinsic 1 - 100 keV luminosity, estimated from the observed scattered X-rays, lies between $10^{43}\,\text{erg}\,\text{s}^{-1}$ and $10^{44}\,\text{erg}\,\text{s}^{-1}$. The largest uncertainty in the X-ray luminosity is the fraction of X-rays that is scattered into our line of sight.  \citet{Pier-1994} give a comparison of different methods to determine this fraction, from which we derive an intrinsic luminosity $L_{1-100\,\text{keV}} = 10^{43.5}\,\text{erg}\,\text{s}^{-1}$, and a spectral slope $F_{\nu} \sim \nu^{-1}$ (or photon index $\Gamma = 2$). 

Further out from the obscuring torus, the AGN is surrounded by a ring of molecular gas with a radius of $\sim 1''-5''$ (72-360 pc), also called the circumnuclear disk (CND). The CND is observed in CO and H$_{2}$ \citep{Schinnerer-2000, Mueller-2009}, and observations of SiO, HCN and CN suggest an X-ray driven chemistry \citep{Lepp-1996, Usero-2004, Garcia-Burillo-2010}. 

Inside this shell, the near-infrared integral field spectrograph SINFONI provided 0.075$''$ resolution observations of H$_{2}$~1-0~S(1), which reveal two streamers within 0$''$.4 (30 pc) \citep{Mueller-2009}. The southern streamer lies between us and the AGN center. The SINFONI observations furthermore reveal the presence of the northern streamer (0$''$.4 or 30 pc north of the AGN center), which may fuel the central black hole.

A ring of H$_{2}$O masers, with a radius of $\sim$ 20 mas (1.5 pc), is centered at the AGN \citep{Gallimore-1996}. The water masers trace densities $n_{H} > 10^{7}\,\text{cm}^{-3}$, and could well locate the inner edge of a molecular torus. The dust in the torus, on the other hand, is best observed in the IR continuum. The NGC\,1068 nucleus has been observed several times with the mid-infrared interferometer MIDI in order to determine the dust distribution.  \citet{Lopez-2014} present the most recent observations, which resolve the central 1-10 pc of the AGN. The observations are best fit with 3 gray bodies with a Gaussian intensity distribution: a $\sim 1.4\,\text{pc} \times 0.5\,\text{pc}$ component of 700 K, a  $\sim 3\,\text{pc} \times 2\,\text{pc}$ component of $\sim 300\,\text{K}$, and a component offset to the west (by $1.7\,\text{pc}$) and north (by $5.6\,\text{pc}$) with a projected size of $\sim 14\,\text{pc} \times 3.5\,\text{pc}$ and a temperature of $\sim 360\,\text{K}$.

\section{PACS Observations and Data Reduction}

\label{sec:observations}
The CO ladder of the NGC\,1068 nucleus has been observed up to $\text{J}=30$ with the Photo detector Array Camera and Spectrometer (PACS) on board the Herschel Space Observatory, and is described in HD12. Here we present a deeper observation taken around $\lambda = 65.94\,\upmu$m, where we expect the CO(40-39) line with a rest wavelength $\lambda = 65.69\, \upmu$m at a system velocity of $v_{\text{LSR}} = 1125\,\text{km}\,\text{s}^{-1}$. The observations were taken as part of our GT Key Project SHINING, the OBSIDs are listed in Table \ref{tab:full-scan}. Among the high J CO lines, CO(40-39) is particularly suitable because the spectra of bright galaxies have little contamination from other lines around $\lambda = 65.69\, \upmu$m. Moreover, the line traces gas with densities $> 10^{6}\,\text{cm}^{-3}$ and temperatures $> 500\,\text{K}$ (Table \ref{tab:j40}), and distinguishes a highly excited component from the gas in the CND and streamers.

\begin{deluxetable}{ccc}
 \tablecaption{Observed wavelengths}
 
 \tablehead{\colhead{OBSID} & \colhead{Blue band} & \colhead{Red band} \\ 
\colhead{} & \colhead{($\upmu$m )} & \colhead{($\upmu$m )} } 
 
\startdata 
 
1342203129 & 65.0-70.5 & 130.2-141.2 \\
1342203128 & 58.5-66.0 & 117.4-132.0 \\
1342203127 & 52.0-59.5 & 103.8-119.4 \\
1342203126 & 94.6-98.0 & 188.6-196.0 \\
1342203125 & 91.0-95.1 & 181.4-190.4 \\
1342203124 & 87.3-91.5 & 174.0-183.0 \\
1342203123 & 83.2-87.7 & 166.0-176.0 \\
1342203122 & 78.9-83.7 & 157.0-167.7 \\
1342203121 & 74.4-79.4 & 148.0-159.0 \\
1342203120 & 70.0-75.0 & 139.6-150.0 \\
1342259623 & 65.4-66.5 & \\
1342259624 & 65.4-66.5 & \\
\enddata
\tablecomments{OBSIDs and covered wavelength of the full Herschel-PACS scan of NGC\,1068. All observations for the full spectrum have been performed on observational day (OD) 460. The last two OBSIDs are from the deeper observations on CO(40-39). }
 \label{tab:full-scan}
\end{deluxetable}

\begin{deluxetable}{ccccc}
\tablecaption{Details of the CO(40-39) line}

\tablehead{\colhead{transition} & \colhead{rest wavelength} & \colhead{upper limit} & \colhead{$n_{crit}$} & \colhead{E$_{upper}$/$K_{B}$} \\ 
\colhead{} & \colhead{($\upmu$m)} & \colhead{($10^{-17}\, \text{ W m}^{-2}$)} & \colhead{($\text{cm}^{-3}$)} & \colhead{(K)} } 

\startdata
CO(40-39)  & 65.69 & 2 & $10^{7}$ & $4 \times 10^{3}$  \\
\enddata

\label{tab:j40}
\end{deluxetable}

We used HIPE 11 for the data reduction. The previous data were analyzed with HIPE 6 (HD12), so we re-reduced some of the $\text{J} \leq 30$ lines with HIPE 11.
The high-J lines detected in NGC\,1068 are all consistent with arising in the central spatial pixel which covers the central $\sim (9''.4)^{2}\,\text{or }\, (700\,\text{pc})^{2}$ of NGC\,1068.
Therefore, the data reduction discussed here applies to the central spaxel only. The spectra were reduced with the standard pipeline for a ``chopped line scan and short range scan with background normalization'', including the point source correction factor \citep{PACS-2010}, and setting the up-sample parameter to 1 and the over-sample parameter to 2. We found that the CO(30-29) flux should be slightly modified, we now measure $(2 \pm 1) \times 10^{-17}\,\text{W}\,\text{m}^{-2} $, instead of $(4.2 \pm 1.9) \times 10^{-17}\,\text{W}\,\text{m}^{-2} $.  There is one other small modification we make to the error bars published in HD12: the error in CO(22-21) is larger than 30\% of the total flux. This line is observed at the edges of two adjacent spectral scans, and therefore the uncertainty in the flux is larger than in other lines. The modified line flux is $(4^{+4}_{-1.4}) \times 10^{-17}\, \text{W}\,\text{m}^{-2} $.

\begin{figure}
   \centering
   \includegraphics[width=8cm]{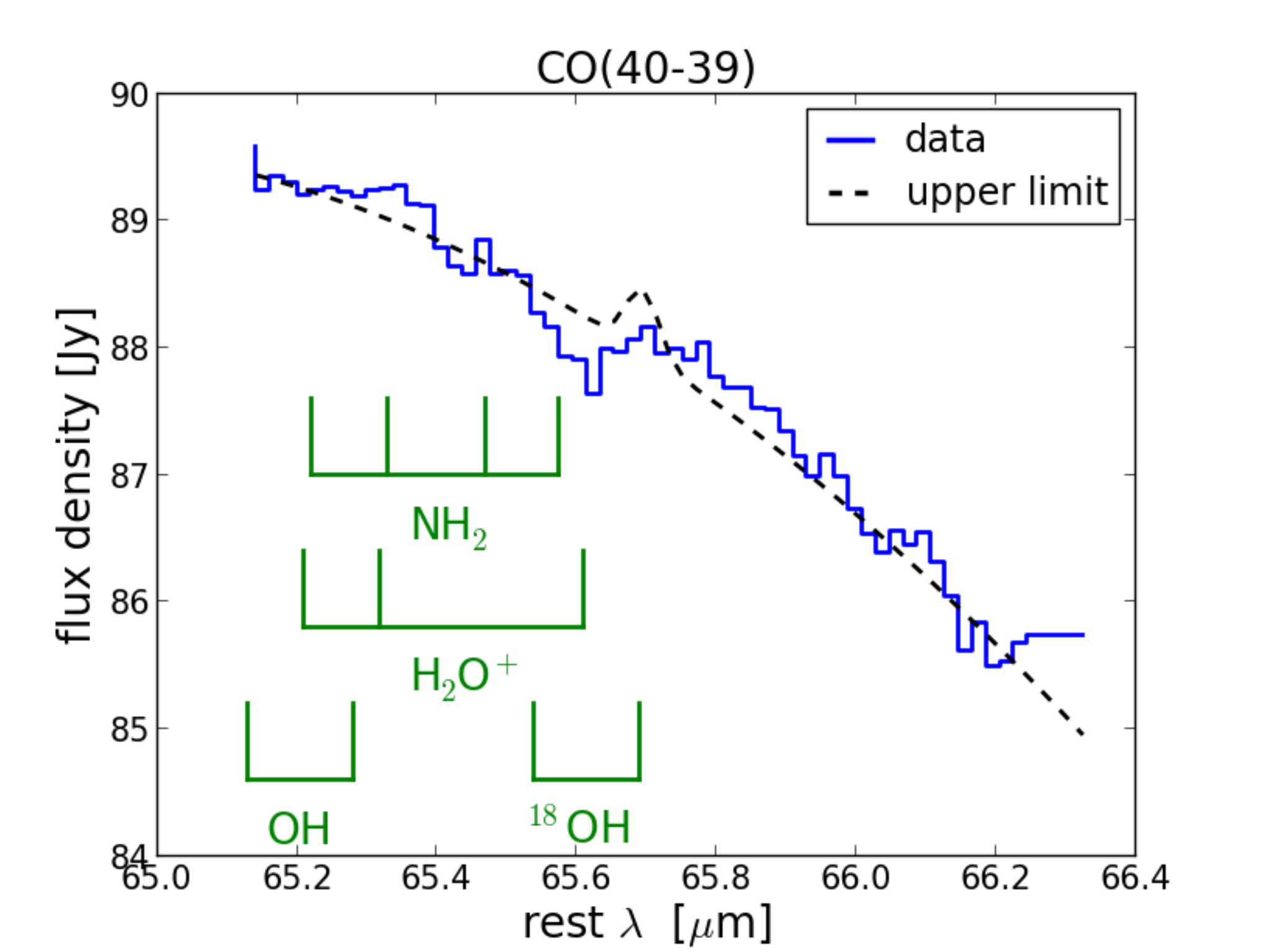}
   \caption{
   Continuum fit and $3 \sigma$ upper limit: A Gauss profile with a flux of $2 \times 10^{-17}\, \text{ W m}^{-2}$ and a FWHM of 250 km/s illustrates the $3 \sigma$ upper limit of the CO(40-39) line, at the system velocity of NGC 1068 and a rest wavelength of $65.69\, \upmu$m. The feature around 65.6 $\upmu$m is considered an artifact. }
 \label{fig:linefit}%
    \end{figure}

The CO(40-39) line observations have been reduced with the same pipeline and parameter setting as described above.  Figure \ref{fig:linefit} shows the region of the CO(40-39) line in the rest frame with a spectral resolution of $0.04\,\upmu$m. There is no clear emission at $\lambda = 65.69\, \upmu$m, but there could be an absorption feature around $\lambda = 65.6\, \upmu$m. Potential absorbers are $^{18}$OH $\Pi_{3/2} - \Pi_{3/2} \frac{9^{+}}{2^{\phantom{+} }} - \frac{7^{-}}{2^{\phantom{-} }}$ at $65.69\,\upmu$m, the CO(40-39) line itself, $\text{H}_{2}\text{O}^{+}$ $3_{31}-2_{20}\, 7/2-5/2$ at $65.61\, \upmu$m, and NH$_{2}$ at $65.57\, \upmu$m. From our PACS observations of other galaxies at both $65\, \upmu$m and $120\, \upmu$m the line flux ratio of $^{16}$OH to $^{18}$OH is always found to be equal to or greater than 10 \citep{Gonzalez-2012, Gonzalez-2014}. Because there is no $^{16}$OH absorption visible in Figure \ref{fig:linefit} at $65.28\,\upmu$m, $^{18}$OH can not be responsible for the absorption 
feature at  $\lambda = 65.6\, \upmu$m. Moreover, if identified with $^{18}$OH or CO(40-39), the line would be blue-shifted by $400\,\text{km}\,\text{s}^{-1}$ with respect to the host galaxy, while the highest velocity  features in the AGN nucleus are the water masers, with $\sim 300\,\text{km}\,\text{s}^{-1}$ \citep{Gallimore-1996}. 

\begin{figure}
 \includegraphics[scale=0.45]{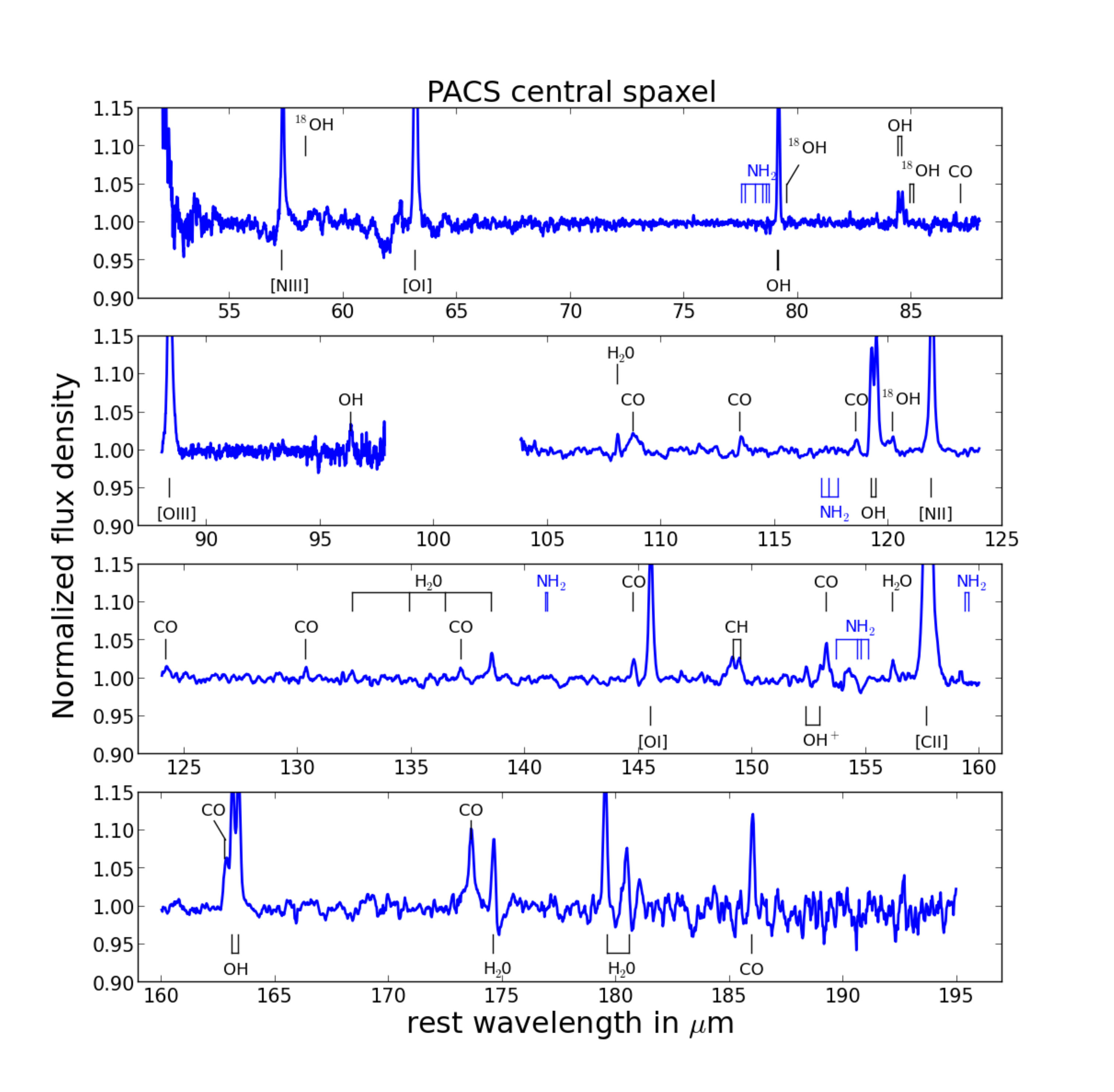}
 \caption{The full PACS spectrum of the central spaxel. The wavelength of the brightest NH$_{2}$ lines are indicated in the spectrum, to show that they are very faint (if detected at all). At 119 $\upmu$m, the flux ratio $^{16}\text{OH}$/$^{18}\text{OH} \approx 10$, so if no $^{16}\text{OH}$ is observed at $65.28\,\upmu$m, the $^{18}\text{OH}$ line can not cause the absorption at $65.6\, \upmu$m.  61 to 63$\, \upmu$m is a known region with artifacts in the relative spectral response function. }
 \label{fig:pacs-spectrum}
\end{figure}

Absorption by NH$_{2}$ is unlikely because it should have several features of comparable intensity as indicated in Figure \ref{fig:linefit}, and these are not observed. Moreover, the other NH$_{2}$ lines in NGC 1068 are faint. See Figure \ref{fig:pacs-spectrum} for the full (55 to 200 $\upmu$m) PACS spectrum from the central spaxel of NGC\,1068.   $\text{H}_{2}\text{O}^{+}$ absorption cannot be excluded completely, with the $65.61\,\upmu$m line the strongest of the three lines in Figure \ref{fig:linefit}, but it would be an exception in an otherwise emission dominated spectrum. A detailed analysis of the PACS spectra of NGC\,1068 and other objects will be given in Fischer et al. (in preparation).

Therefore, we assume that the feature around $\lambda = 65.6\, \upmu$m is an artifact, and treat it as noise.  This results in a noise level of 0.12 Jy, and a 3$\sigma$ upper limit on CO(40-39) of $2 \times 10^{-17}\,\text{W}\,\text{m}^{-2}$, for a line width of $250\,\text{km}\,\text{s}^{-1}$ (HD12).  Figure \ref{fig:linefit} shows the upper limit superposed on the spectrum. If the feature is masked out during the fit, the 3$\sigma$ upper limit becomes $1 \times 10^{-17}\,\text{W}\,\text{m}^{-2}$, and the absorbed flux is $4 \times 10^{-17}\,\text{W}\,\text{m}^{-2}$. The feature could have absorbed the blue part of the CO(40-39) line, but it is hard to quantify by how much. Because there is no red wing visible, the artifact would have to be coincidently just equal to that of a possible CO(40-39) emission line in order to mask it. We find this scenario unlikely and use the aforementioned upper limit of $2 \times 10^{-17}\,\text{W}\,\text{m}^{-2}$ for our analysis. The CO SLED with the upper 
limit and the modifications to HD12 is given in Figure \ref{fig:fits}.

\section{PDR, XDR, and shock models}
\label{sec:xdr-fit}
With the slightly modified CO(30-29) value and the new CO(40-39) upper limit we repeated our PDR, XDR and shock fits, with the same models as in HD12. For comparison we briefly summarize the main results from HD12 here. The observed CO SLED was analyzed with Large Velocity Gradient- (LVG), shock-, PDR-, and XDR-models. Only the PACS data were fitted; the $3\,<\,J\,<\,14$ CO lines which trace the colder gas were analysed by \citet{Spinoglio-2012}. The PACS SLED is best explained with two components: a moderately excited (ME) component with $T_{kin} \approx 169\,$K, $n_{H_{2}} \approx 10^{5.6}\, \text{cm}^{-3}$ and a total mass $M_{H_{2}} \approx 10^{6.7}\, \Msun$, and a highly excited (HE) component with $T_{kin} \approx 571\,$K,  $n_{H_{2}} \approx 10^{6.4}\, \text{cm}^{-3}$ and a total mass $M_{H_{2}} \approx 10^{5.7}\, \Msun$. The ME component can arise from X-ray- or shock-heated gas in the CND, while the HE component can arise from either the CND or the northern streamer. In order to determine 
whether 
the HE component arises from the CND or the northern streamer, we followed H12 and compared the PACS CO 
line shifts with the SINFONI observations of H$_{2}$ 1-0 S(1). The PACS HE lines are blueshifted by $59 \pm20 \, \text{km s}^{-1}$, which best matches the $\sim 25\, \text{km s}^{-1}$ blueshift of the northern streamer.

To fit our new results, we use the XDR grid from \citet{Meijerink-2007}. This grid contains the CO line emission for gas with densities $n_{H} = 10^{4} - 10^{6.5}$ cm$^{-3}$ and X-ray fluxes $F_{X} = 1.6 - 160\, \text{erg}\, \text{cm}^{-2}\, \text{s}^{-1}$. In HD12, the HE component was best fit with $F_{X} = 160\, \text{erg}\, \text{cm}^{-2}\, \text{s}^{-1}$, i.e. the maximum value in the grid. This time we used an extended grid with  $F_{X} = 9 - 510\, \text{erg}\, \text{cm}^{-2}\, \text{s}^{-1}$ (R. Meijerink, private communication).

\begin{figure}
 \includegraphics[scale=0.4]{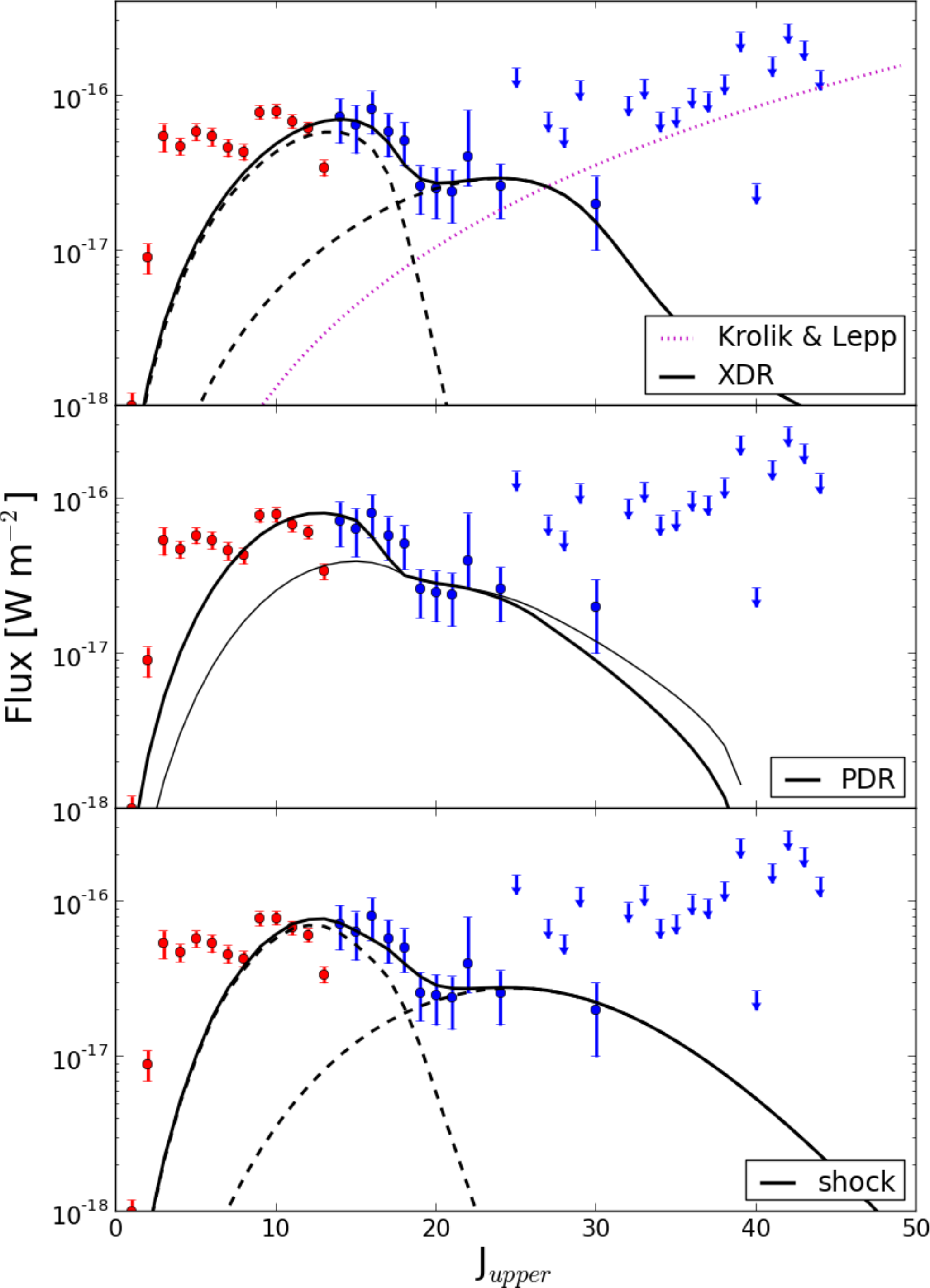}
 \caption{ Best XDR- PDR- and shock-fits to the CO SLED.  Red points indicate Plateau de Bure Interferometer data ($\text{J from } 1-3 $, with beam size ranging from 0''.6 to 2''.5; \citet{Krips-2011}) and Spire data ($\text{J from } 4-13 $, with beam sizes ranging from 17'' to 42''; \citet{Spinoglio-2012}), blue points represent PACS data  ($\text{J} > 13 $, with beam size ranging from 11'' to 14''; HD12). Only the PACS observations were used for the fits. Upper panel: A two component XDR fit with an extended grid of \citet{Meijerink-2007}, and expected CO emission from an X-ray heated molecular torus with $N_{H} = 10^{24}\,\text{cm}^{-2}$ and $f_{cov}f_{x44} = 0.08$ (KL89).
 Central panel: PDR fit to the full PACS SLED (thick line), and a PDR fit to the $J \geq 19$ observations (thin line). 
 Lower panel: a two component shock-fit.
  Model parameters of all fits are discussed in the text and in Table \ref{tab:fits}. }
 \label{fig:fits}
\end{figure}

Table \ref{tab:fits} and Figure \ref{fig:fits} show the parameters of the best XDR-, PDR- and shock-fits. Only detected lines were used during the fit, but the best fits were in all cases consistent with the upper limits of the high-J CO lines.  The XDR fit of the HE component is slightly different from HD12, due primarily to the lower value for CO(30-29) and to the extension of the grid: the best fit now has an input X-ray flux $F_{X} = 510\, \text{erg}\, \text{cm}^{-2}\, \text{s}^{-1}$, which is again the high end point of the grid. We tested whether an even higher X-ray input flux would better fit the HE component, using a different XDR grid (described in Section \ref{sec:xdr-grid}), which contains X-ray fluxes up to $F_{X} = 10^{6}\, \text{erg}\, \text{cm}^{-2}\, \text{s}^{-1}$. For a thermal X-ray input spectrum like in the grid from Meijerink, the CO SLEDs from the 2 codes agree within 20\% for J $< 30$. 
In this model an input flux of $F_{X} = 10^{3}\, \text{erg}\, \text{cm}^{-2}\, \text{s}^{-1}$ could still fit the HE component, but is no improvement over the fit with $F_{X} = 510\, \text{erg}\, \text{cm}^{-2}\, \text{s}^{-1}$.

Our shock modeling uses a two component fit: the ME C-shock model is from Flower \& Pineau Des For\^{e}ts (2010), and the HE C-shock model is from Kaufman \& Neufeld (1996). The lower value of the CO(30-29) line has a small effect on the best fit of a shock model to the HE component. The HE component can now be fit with a 30 km/s velocity, instead of 40 km/s. The effect on the PDR modeling is small (using the PDR model from \citet{Meijerink-2007}). A PDR cannot be ruled out,  but it is hard to match the $\text{J} = 30$ line without overproducing the lower J lines.  A single PDR component only marginally fits the entire SLED (see Figure \ref{fig:fits}). Because the kinematics of the lines suggest two components, we discard this option. The PDR fit to the HE component ($J \geq 19$) alone is slightly improved with the lower CO(30-29) flux, but not as good as an XDR or shock fit.

The upper panel in Figure \ref{fig:fits} also shows the expected CO emission from a pc-scale molecular torus with $N_{H} = 10^{24}\,\text{cm}^{-2}$ and  $f_{cov}f_{x44} = 0.08$, as predicted by KL89,  where $f_{cov} = 0.25$ is the surface covering factor of the torus and $F_{X} = F_{x44} \times 10^{44}\, \text{erg}\, \text{cm}^{-2}\, \text{s}^{-1}$. This is further discussed in section \ref{sec:slab}.

We conclude that the revised value for CO(30-29) and the new upper limit on CO(40-39) do not significantly change the results from HD12. They found that the ME component can arise from X-ray- or shock-heated gas in the CND, while X-ray heated gas in the northern streamer is the preferred explanation for the HE component, based on the fits and the kinematics.

\begin{deluxetable}{ccccc}
 \tablecaption{Heating mechanisms}

\tablehead{ \colhead{} & \colhead{ME} & \colhead{HE} & \colhead{Full} & \colhead{$\chi^{2}$} \\ 
 } 
 
\startdata
 
XDR & $n_{H} = 10^{5.25}\, \text{cm}^{-3}$ & $n_{H} = 10^{6.5}\, \text{cm}^{-3}$ & ... & \\
    & $F_{X} = 9\, \text{erg}\, \text{cm}^{-2}\, \text{s}^{-1}$ & $F_{X} = 510\, \text{erg}\, \text{cm}^{-2}\, \text{s}^{-1}$ & ... & 1.23\\
    & $A \approx (108\, \text{pc})^{2}$  & $A \approx (20\, \text{pc})^{2}$  &  ... & \\
    \hline
PDR & ... & $n_{H} = 10^{6.5}\, \text{cm}^{-3}$ & $n_{H} = 10^{6}\, \text{cm}^{-3}$ &\\
    & ... & $G_{0} = 10^{5}$ & $G_{0} = 10^{5}$ &  2.16\\
    & ... & $L_{FUV} \approx 3 \times 10^{9} L_{\sun}$ & $L_{FUV} \approx 10^{10} L_{\sun}$ &\\
    \hline
C-shock & $n_{0} = 2 \times 10^{5}\, \text{cm}^{-3}$ & $n_{0} = 10^{6}\, \text{cm}^{-3}$ & ...&  \\
	& $v = 20\, \text{km}\, \text{s}^{-1}$ & $v = 30\, \text{km}\, \text{s}^{-1}$ & ... & 1.88 \\
	& $A \approx (150\, \text{pc})^{2}$ & $A \approx (19\, \text{pc})^{2}$ & ... & \\
\enddata
	
\tablecomments{Parameters for the best fits with XDR-, PDR-, and shock-models. The XDR and PDR models are from \citet{Meijerink-2007}, with the XDR grid extended to $F_{X} = 510\, \text{erg}\, \text{cm}^{-2}\, \text{s}^{-1}$.  Two models  were employed for the shock-fit: the ME component is modeled with \citet{Flower-2010}, and the HE component with \citet{Kaufman-1996} (like in HD12). $A$ is the illuminated area. The $\chi^{2}$ values apply to the 2 component XDR- and shock- fit, and to the full PDR fit. }
\label{tab:fits}
\end{deluxetable}

\section{XDR model of the NGC\,1068 nucleus}
\label{sec:model}
 In the previous section we have attributed the HE component of the high-J CO SLED to the northern streamer, at about 30 pc from the nucleus. If this is correct then we have not detected CO emission from the putative torus that is required by AGN unification models. In this section we want to address the following question: what does the upper limit on CO (40-39) tell us about the existence or non-existence of a torus? If there is such a torus, is it reconcilable with the upper limit, and what can be learned about its physical parameters (like size, density, distance from the center)?

In the following sections we model the molecular torus as a semi-infinite slab or as a system of such slabs, with strong irradiation from X-rays. The intrinsic X-ray luminosity $L_{1-100\,\text{keV}} = 10^{43.5}\,\text{erg}\,\text{s}^{-1}$ indicates input fluxes up to $F_{X} = 10^{6}\, \text{erg}\, \text{cm}^{-2}\, \text{s}^{-1}$ are needed to account for gas as close as $\sim 0.5$ parsec from the AGN. Different from the previous section (and from HD12) we here use our own XDR code as described in \citet{Bruderer-2012} (including benchmark tests), with updates in \citet{Bruderer-2013}. This code, originally developed for the analysis of proto-planetary disks, was extended to the higher X-ray input fluxes we need here.

\subsection{XDR grid}
\label{sec:xdr-grid}

In the model, X-rays with photon energies between 1 keV and 100 keV deposit energy in the gas, which has two effects: they heat the dust and drive ionization reactions. The code iterates on radiative, thermal and chemical processes, until all are balanced. From this balanced situation, the gas temperature and CO rotational line emission is written out as a function of distance and column density into the slab. 

In a semi-infinite slab model, the observer is located on the same side as the X-ray source. X-rays enter the semi-infinite slab from one side and deposit their energy in the gas. CO is emitted in all directions, but only the CO emission towards the observer is calculated. CO can become optically thick between the point of emission and the observer (we did not assume the gas to have a large velocity gradient), in this case the attenuated CO emission is given. NGC 1068 is however a type 2 AGN, so the gas is located between the observer and the X-ray source. To check whether the location of the observer introduces a large uncertainty in our results, we extracted the optical depths as a function of column density from the XDR code, for all CO lines. With these optical depths, the unattenuated CO emission was calculated as a function of column density, as well as the attenuated CO emission towards an observer on the opposite side of the slab. Indeed, for some combinations of density and X-ray 
flux, the CO becomes very optically thick and the observed CO emission differs by more than an order of magnitude from one side of the slab to the other. For the parameters that are relevant for this paper however,  the difference is at most 30\% in CO line flux at column densities $< 10^{25}\,\text{cm}^{-2}$. Another potential issue is geometrical dilution, this is discussed further in Section \ref{sec:clumpy}.

The expected CO emission is calculated for a grid of semi-infinite slabs with densities n $\in$ [$10^{2}$, $10^{2.5}$ ... $10^{9}$] cm$^{-3}$ and F$_{X}$ $\in$ [$10^{-4}$,$10^{-3}$ ... $10^{6}$] $\text{erg}\,\text{cm}^{-2}\,\text{s}^{-1}$.
The input SED is a power law with the shape $F_{\nu} \propto \nu^{-1}$, which best matches the observed XDR spectrum of NGC 1068 \citep{Pier-1994}. We also ran the grid for a softer intrinsic spectrum (spectral index of -2 instead of -1). In this case, the X-ray absorption and CO emission takes place within smaller column densities than for the hard X-ray spectrum. 

Line absorption by dust particles is not taken into account, because this depends strongly on geometry and viewing angle. Since we do not know the geometry of gas structures in the central few parsec of NGC\,1068, we rather assume no line absorption by dust. This can be done safely up to a column $N_{H} = 8.5 \times 10^{23} \text{cm}^{-2}$, where the dust becomes optically thick at 66 $\upmu$m \citep{Draine-2003a}. At higher column densities, the results from our model are thus upper limits to the line fluxes (see Section \ref{sec:dust}).

\begin{figure}
 \includegraphics[scale=0.3]{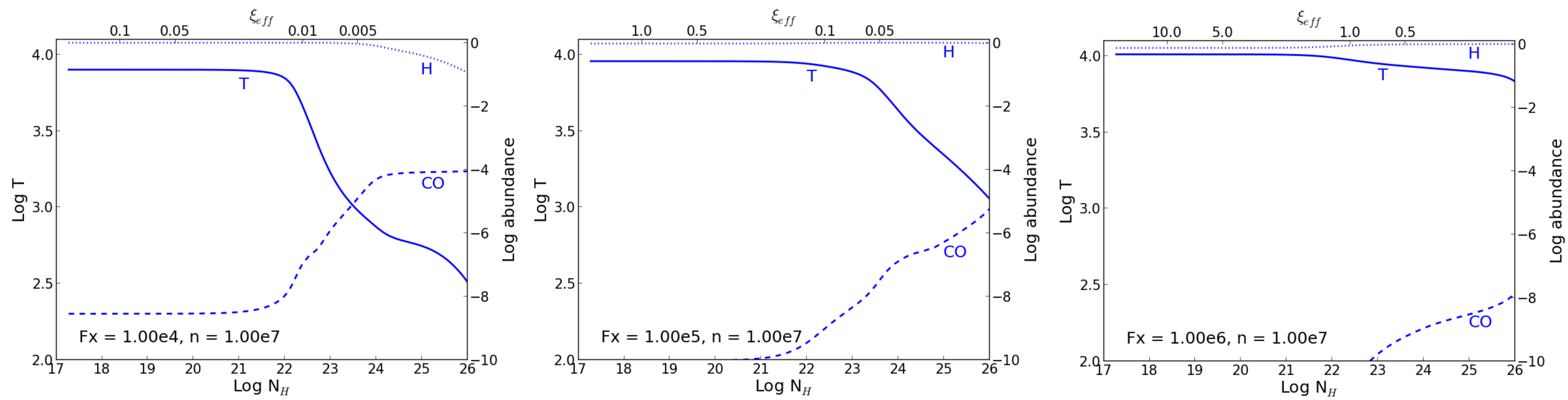}
 \caption{The gas temperature, effective ionization parameter, and CO- and H- abundances are plotted as function of column density into the slab. The density is $10^{7}\,\text{cm}^{-3}$ in all the figures. Left: $F_{X} = 10^{4}\,\text{erg}\,\text{cm}^{-2}\,\text{s}^{-1}$, the gas has become molecular around $N_{H} = 10^{24}\,\text{cm}^{-2}$. Middle: $F_{X} = 10^{5}\,\text{erg}\,\text{cm}^{-2}\,\text{s}^{-1}$, the gas becomes slowly molecular and produces most CO(40-39) emission around $N_{H} = 10^{25}\,\text{cm}^{-2}$, with $\xi_{eff} \approx 0.02$ (see Figure \ref{fig:one_slab}). Right: $F_{X} = 10^{6}\,\text{erg}\,\text{cm}^{-2}\,\text{s}^{-1}$, even at  $N_{H} = 10^{26}\,\text{cm}^{-2}$, there is too little CO to produce observable CO(40-39) emission. }
 \label{fig:4axes}
\end{figure}

In all grid points, the line emission is calculated up to a column of $N_{H} = 10^{26}\, \text{cm}^{-2}$ into the slab. High-J CO emission arises from a certain combination of density n, X-ray flux F$_{\text{x}}$, and column density, best represented by the effective ionization parameter $\xi_{eff}$. $\xi_{eff}$ is calculated from formula 12 in \citet{Maloney-1996}; $\xi_{eff} = 4\pi \frac{\text{F}_{\text{x}}}{\text{n}\,N_{22}^{\phi + 1}}$, with $N_{22}$ the column density in units of $10^{22}\,\text{cm}^{-2}$, $\phi = (\alpha -1)/\gamma$, in which $\alpha = -1$ is the spectral index, and $\gamma = 8/3$ is a parameter that reflects that we are interested in the high energy X-rays (E $> 1$ keV). The high-J CO lines (around J = 40) are well produced for an effective 
ionization parameter $\xi_{eff} \approx 0.02$. $\xi_{eff}$ is calculated for 3 grid points with $n = 10^{7}\,\text{cm}^{-3}$, and $F_{X} = 10^{4}, 10^{5}\,\text{or}\, 10^{6}\,\text{erg}\,\text{cm}^{-2}\, \text{s}^{-1}$, and plotted in Figure \ref{fig:4axes}, together with the gas temperature and the H- and CO- abundances. The central grid point produces most CO(40-39) emission: at higher $F_{X}$, the gas remains hot and the CO abundance low, while at lower $F_{X}$ the gas quickly becomes molecular and cools rapidly. 

Although H$_{2}$ is the main collision partner of CO in a molecular gas, atomic hydrogen also contributes to collisional excitation of CO if the gas is partly atomic or ionized. Because collision rates between H and CO are currently still uncertain, they are approximated by the H$_{2}$ collision rate, scaled with the reduced mass of the colliding particles. The next Section treats the CO SLED that results from our model.

\subsection{An X-ray irradiated slab}
\label{sec:slab}
A model of a torus with a smooth plane parallel slab that is intensively radiated with X-rays was already given by KL89, who concluded that most of the torus cooling happens in (very) high J CO lines (with J up to $\sim 57$). 
A certain column density is however required to produce observable CO fluxes, because the gas has a stratified structure: the first layer, through which the X-rays enter the gas, is fully ionized with temperatures of a few thousand Kelvin. The CO forms only further into the slab, and cools the gas rapidly.

In the following discussion the illuminated slab has a covering factor 0.25 since 20-25\% of all AGN are compton thick \citep{Malizia-2009, Burlon-2011} and its illuminating X-ray flux depends solely on the intrinsic luminosity of NGC 1068 ($10^{43.5}\,\text{erg}\,\text{s}^{-1}$, assuming an isotropically radiating source) and the distance towards the AGN. 
The absolute emitted CO line fluxes of these scenarios are subject to large uncertainties \citep{Rollig-2007}, especially at high energy levels. One of the reasons for the uncertainties is that the main collision partner for CO, H$_{2}$, forms on dust grains. The  H$_{2}$ abundance thus depends strongly on the dust abundance, its composition, and its temperature, which are all uncertain parameters. Moreover, the dust and gas temperatures decouple in the extreme conditions considered here, with T$_{\text{dust}} \sim $ a few 100 K, and T$_{\text{gas}} \sim $ a few 1000 K. So unlike in the study of protoplanetary disks, where absolute line fluxes can be given within a factor of a few \citep{Bruderer-2012}, we adopt an uncertainty of 1 order of magnitude in the modeled line fluxes.

\begin{figure}
 \includegraphics[scale=0.3]{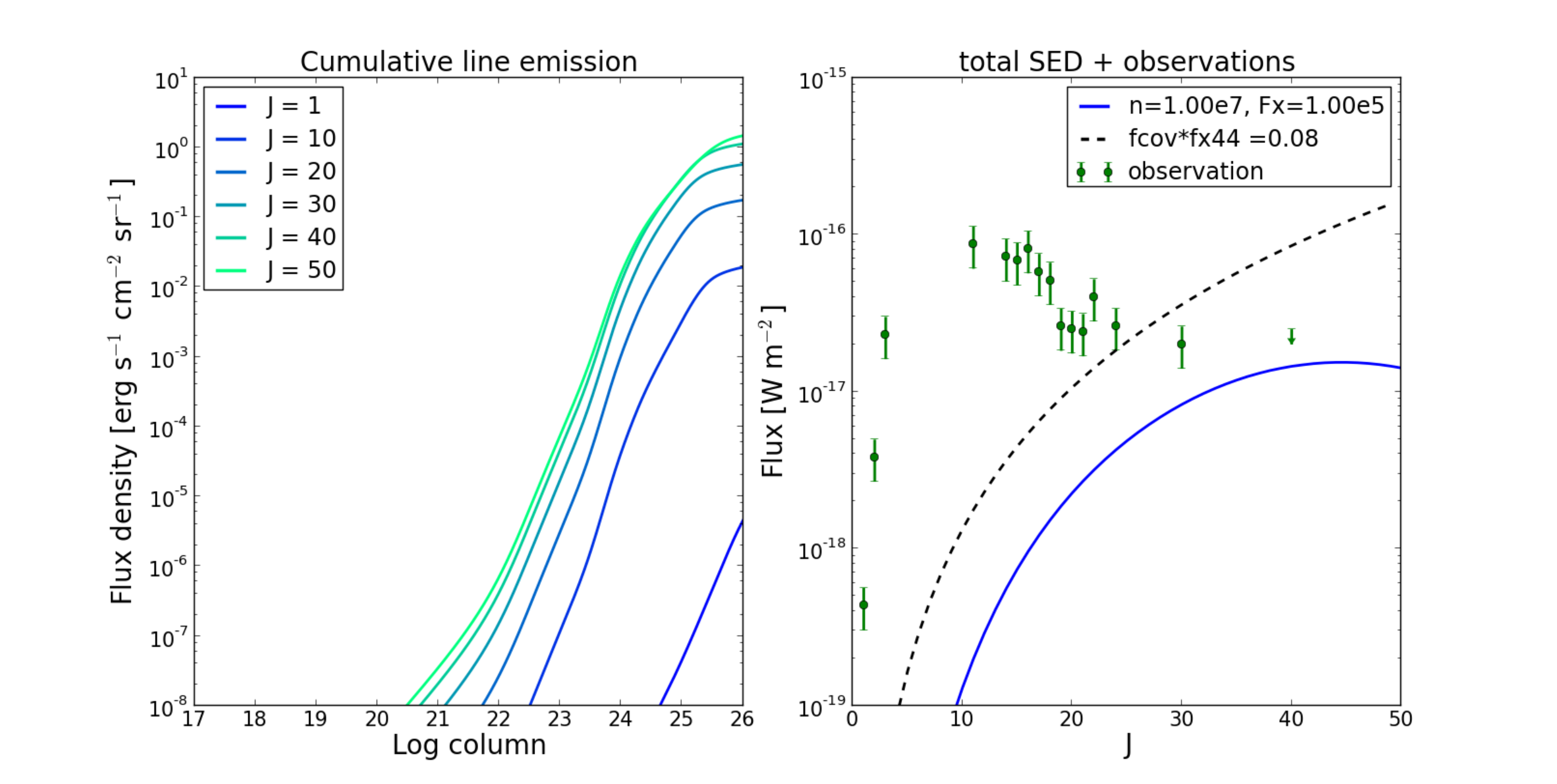}
  \caption{Left: Cumulative line emission from an X-ray irradiated slab with a density $10^{7}$ cm$^{-3}$ and $F_{X} = 10^{5}\,\text{erg}\,\text{cm}^{-2}\,\text{s}^{-1}$. Right: the CO SLED from a $8.3$ pc$^{2}$ sized structure, at a distance of 14.4 Mpc, and with $N_{H} = 10^{25}\,\text{cm}^{-2}$, is plotted together with the prediction by KL89 with $f_{cov}f_{x44} = 0.08$. The difference between the 2 models is a factor 8 for CO(40-39). }
 \label{fig:one_slab}
\end{figure}

A general discussion on what part of parameter space can be constrained by the upper limit is given in Section \ref{sec:discussion}. But first we analyse a simple slab model with parameters based on the water masers described in \citet{Gallimore-1996}. In this model, the gas has a density $n = 10^{7}\, \text{cm}^{-3}$ and an impinging X-ray flux of $10^{5}\, \text{erg}\, \text{cm}^{-2}\, \text{s}^{-1}$, because the water masers imply gas densities of $\geq 10^{7}\,\text{cm}^{-3}$ at 1.5 pc from the nucleus. These are the same parameters as used by KL89, and with an intrinsic X-ray flux $F_{X} = 10^{43.5}\,\text{erg}\,\text{s}^{-1}$, they represent a torus with inner radius of 1.6 pc (our intrinsic luminosity and spectral range differs slightly from KL89). 
Our model prediction of the CO(40-39) flux is a factor $\sim 8$ lower than presented in KL89 -- with $f_{cov}f_{x44} = 0.08$ to account for a covering factor 0.25 -- as can be seen in the right panel of Figure \ref{fig:one_slab}. The reason is that KL89 assume the high-J CO lines to be optically thick up to J $\sim 57$, and therefore the luminosity of the line $L_{J} \propto \nu^{3} \propto \text{J}^{3}$. Instead, we calculate the optical depth as function of column density into the slab, and find that the CO(40-39) line does not become optically thick within $N_{H} = 10^{25}\,\text{cm}^{-2}$. For larger column densities our model converges towards the KL89 prediction (but needs  $N_{H} > 10^{26}\,\text{cm}^{-2}$ to match it).  At $N_{H} = 10^{24}\,\text{cm}^{-2}$, 
as used in KL89, the CO(40-39) flux differs by a factor 100 between the 2 models. It clearly matters at what column density the high-J CO lines become optically thick.

The upper limit corresponds to a column density of $N_{H} = 1.5 \times 10^{25} \text{cm}^{-2}$, which is still similar to the expected obscuring gas column density in NGC 1068 \citep{Bauer-2014}. So although the CO(40-39) line traces the gas we are interested in ($n \sim 10^{7}\,\text{cm}^{-3}$, $N_{H} \sim 10^{25}\,\text{cm}^{-2}$, located within a few pc of the AGN), the observation did not reach the required sensitivity to constrain the gas distribution.

\subsection{A clumpy cloud system}
\label{sec:clumpy}
In the above scenario the gas is distributed smoothly. There are however indications that the dust in a torus is distributed in clouds, and most recent torus models treat a clumpy, rather than a smooth, dust distribution (see \citet{Hoenig-2006} for such a model on NGC 1068). 

\begin{figure}
 \includegraphics[scale=0.25]{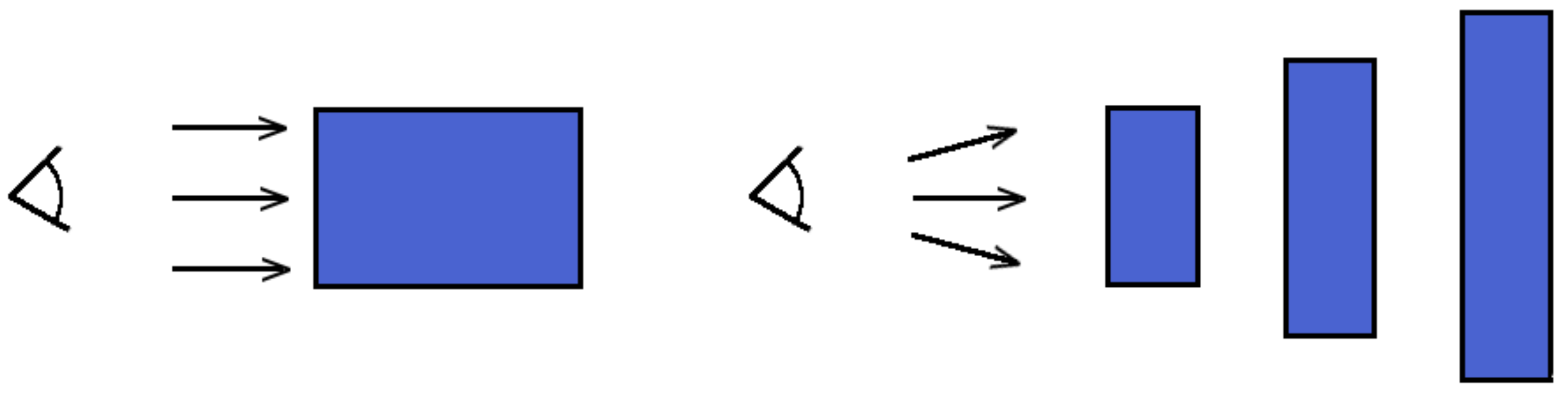}
 \caption{Schematic overview of the torus models. The cartoon on the left represents one slab with a column density $10^{25}\,\text{cm}^{-2}$, the cartoon on the right a 3-layer model with a volume filling factor of 0.5. The column through these 3 slabs is also $10^{25}\,\text{cm}^{-2}$, but the layers have been scaled in height to keep a constant covering factor. The density of all the slabs is constant, while the X-ray flux decreases with distance from the central source to account for geometrical dilution.}
 \label{fig:geometry}
\end{figure}

Assuming that also the gas could be distributed in clouds, we approximate a clumpy torus with 10 plane parallel slabs. The slabs have the same density as in our previous example and the layer closest to the source receives an X-ray input flux of $10^{5}\, \text{erg}\, \text{cm}^{-2}\, \text{s}^{-1}$. All other layers lie at larger distance from the source, and are obscured by the intermediate layers, as shown in Figure \ref{fig:geometry}. The total column density along the line-of-sight to the source is $10^{25}\, \text{cm}^{-2}$, and the illuminated surface of the slabs increases with distance to the source, to keep a covering factor of 0.25. The total amount of gas has therefore increased with respect to the previous example.

The first layer receives the highest X-ray flux, while layers further away from the center receive lower X-ray fluxes due to geometrical dilution and due to X-ray absorption in the intermediate layers. The geometrical dilution is accounted for, by taking the grid point with the X-ray flux that corresponds to the geometrically diluted -- but unattenuated -- X-rays from the central source. In case this flux lies between 2 grid points, the CO SLED is interpolated between these grid points. Each layer thus has a cumulative line distribution attached to it (as shown in the left panel in figure \ref{fig:one_slab}), from which the line emission is extracted between two column densities. For example, the emission of the first layer could thus be extracted from $N_{H} = 0 \rightarrow 10^{24}\,\text{cm}^{-2}$, and that of the second from $N_{H} = 10^{24} \rightarrow 2 \times 10^{24}\,\text{cm}^{-2}$. This accounts for the X-ray attenuation by intermediate layers because the XDR model calculates the X-ray attenuation 
as function of column density. 

\begin{figure}
 \includegraphics[scale=0.3]{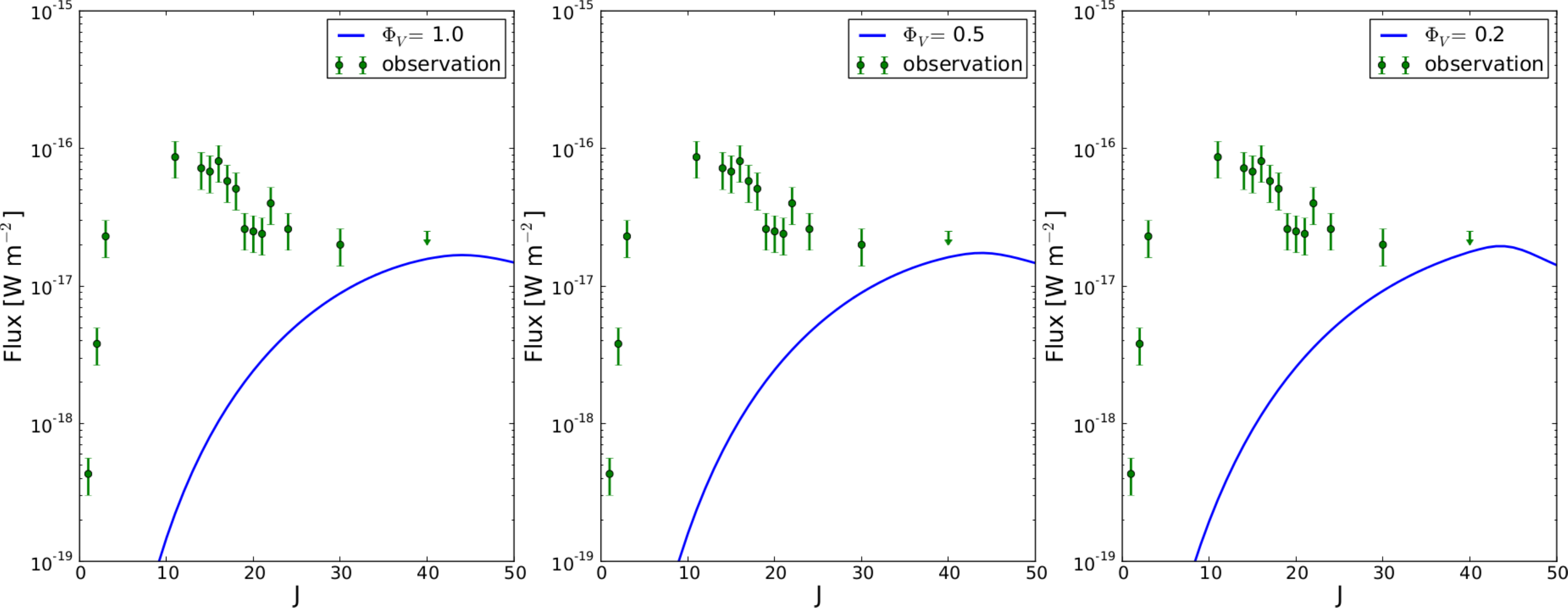}
 \caption{10 plane parallel slabs approximate a torus with $\Phi_{V} = 1$, 0.5, or 0.2. The total column density through the slabs is $10^{25}\,\text{cm}^{-2}$.  Because the increasing slab size compensates the geometrical dilution, the total SLED looks similar in these three cases. }
 \label{fig:3xvff}
\end{figure}

The 10 slabs have column densities of $10^{24}\,\text{cm}^{-2}$. The distance between the slabs is determined by the volume filling factor $\Phi_{V}$: $\Phi_{V} = 0.5$ implies that a 0.03 pc thick layer is followed by a 0.03 pc thick gap. Figure \ref{fig:3xvff} shows the CO SLED with $\Phi_{V} = $ 1, 0.5, and 0.2. The 3 plots look very similar, there is no sign of a lower output flux or change in SLED for lower $\Phi_{V}$. The reason is that the increasing layer surface compensates  the geometrical dilution (both $\propto r^{2}$). Only when $\Phi_{V}$ becomes very small ($\Phi_{V} \sim 0.05$), the outermost layers are located so far from the X-ray source that the CO SLED changes significantly.  We therefore conclude that the volume filling factor is of little importance in our case. 

It is also worth noting that the layered model with $\Phi_{V} = 1$ looks very similar to the single slab model that was presented in Section \ref{sec:slab}. Geometrical dilution is thus not important for the grid point with $n=10^{7}$ cm$^{-3}$, $F_{X} = 10^{5}\,\text{erg}\,\text{cm}^{-2}\,\text{s}^{-1}$ and $N_{H} = 10^{25}\,\text{cm}^{-2}$. The same applies to the other grid points that are used in this paper. 

\subsection{Line absorption by dust}
\label{sec:dust}
Figure \ref{fig:one_slab} showns that column densities of $\sim 10^{25}\,\text{cm}^{-2}$ are needed to match the CO(40-39) upper limit in NGC 1068. Within such high column densities, dust absorbs significant fractions of the line fluxes, and especially affects the high J CO lines.   

\begin{figure}
 \includegraphics[scale=0.4]{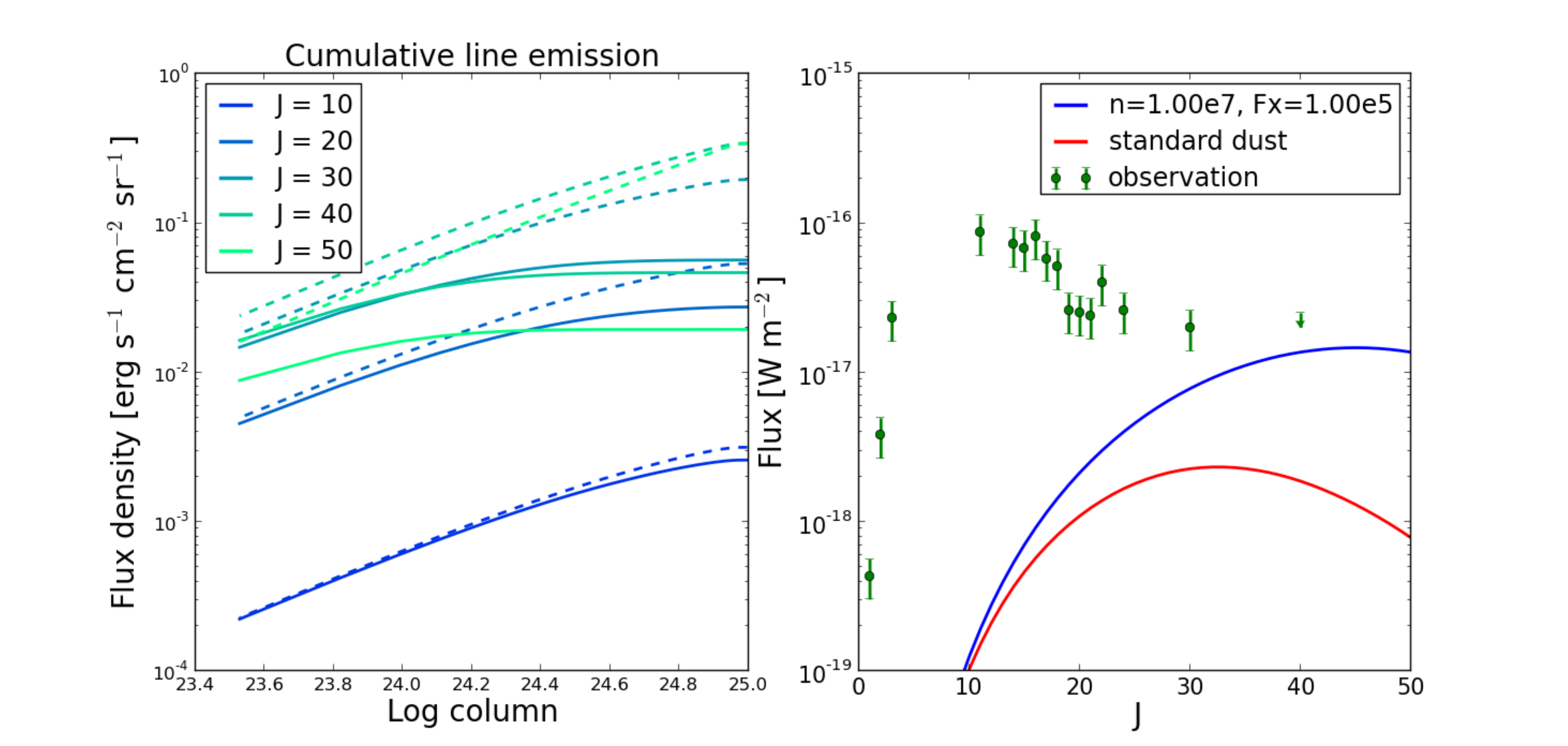}
 \caption{Line absorption by dust in a slab with $n = 10^{7}$ cm$^{-3}$ and $F_{X} = 10^{5}\,\text{erg}\,\text{cm}^{-2}\,\text{s}^{-1}$, calculated for a torus located between the X-ray source and the observer. Left panel: the broken lines indicate unattenuated line emission, while the solid lines indicate attenuated line emission. Right panel: dust attenuates the CO(40-39) emission by a factor 10.}
 \label{fig:dust}
\end{figure}

For simplicity, we assume that the gas and dust follow the same distribution, and that the dust resembles galactic dust. The dust cross sections are taken from \citet{Draine-2003a} and follow $\sigma [\text{cm}^{-2} \text{per H-atom}] \propto \lambda^{-2}$ at the relevant wavelengths \footnote{Actual cross sections for Milky Way, R\_V= 3.1 are found on: \url{http://www.astro.princeton.edu/~draine/dust/dustmix.html}}. 
In this case the position of the observer with respect to the slab does matter for the final SLED. The right panel in Figure \ref{fig:dust} shows the CO SLED when the torus is located between the X-ray source and the observer, taking line absorption by dust into account. The modeled CO(40-39) emission decreased by a factor 10.  Because we do not know the real dust distribution in NGC 1068, we do not include line absorption by dust in the remainder of this paper, but keep in mind that it can attenuate the CO(40-39) flux by a factor 10.

\section{Results and Discussion}
\label{sec:discussion}
A smooth molecular torus (with $n = 10^{7}\,\text{cm}^{-3}$, at 1.6 pc from a $L_{1-100\,\text{keV}} = 10^{43.5}\,\text{erg}\,\text{s}^{-1}$ source, and with column density $N_{H} = 1.5\times 10^{25}\,\text{cm}^{-2}$) could have been observable in CO(40-39) within the sensitivity of the PACS observation. But the uncertainty in the model at these high J lines is an order of magnitude. Moreover the dust could attenuate the CO(40-39) line flux by a factor 10, so this scenario is not strongly constrained by the upper limit.

\begin{figure}
 \includegraphics[scale=0.35]{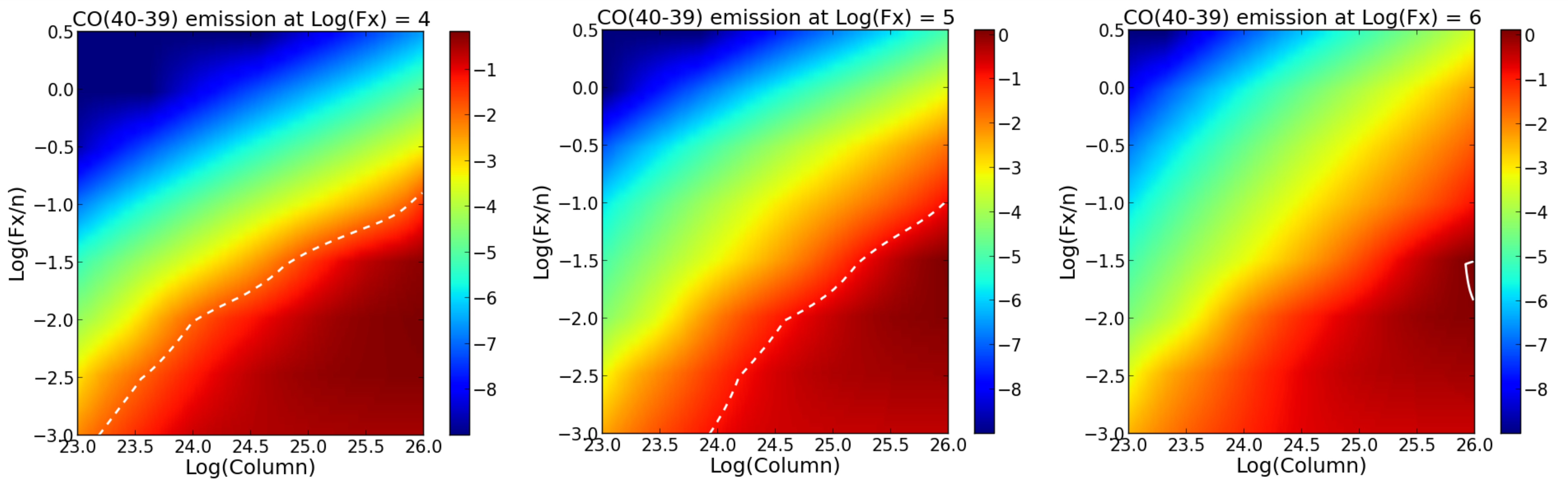}
 \caption{ Cumulative CO(40-39) emission in $\text{erg}\,\text{s}^{-1}\,\text{cm}^{-2}\,\text{sr}^{-1}$. The white broken line indicates where an object with area $4\pi r^{2}$, with $r$ = 5 pc, 1.6 pc, or 0.5 pc respectively (see Table \ref{tab:gas-mass}), matches the upper limit of $2 \times 10^{-17}\,\text{W}\,\text{m}^{-2}$. Taking $N_{H} = 10^{25}\,\text{cm}^{-2}$ and $\text{log}(\frac{F_{X}}{n}) = -2$, the surface covering factor that is needed to match the upper limit is given in Table \ref{tab:gas-mass}. 
 }
 \label{fig:j40-emission}
\end{figure}

Can we exclude anything else? Is there any grid point where the CO emission exceeds the upper limit by more than an order of magnitude? Figure \ref{fig:j40-emission} shows the cumulative CO(40-39) emission as function of column density and $\frac{F_{X}}{n}$.  At $N_{H} > 10^{25}\,\text{cm}^{-2}$ the uncertainty rises above 30\% for a gas located between the X-ray source and the observer. Line absorption by dust is not taken into account. The white, broken, line illustrates where the observed upper limit is matched for an illuminated area of $4\pi r^{2}$, with $r$ = 5 pc, 1.6 pc or 0.5 pc, as given in Table \ref{tab:gas-mass}. 

In all cases, the maximum emission arises at column densities  $ \geq 10^{25}\,\text{cm}^{-2}$, and at $ -2 < \text{log}(\frac{F_{X}}{n}) < -1.5$. So our initial analysis with $n = 10^{7}\,\text{cm}^{-3}$ and $F_{X} = 10^{5}\,\text{erg}\,\text{cm}^{-2}\,\text{s}^{-1}$ already produced CO(40-39) very efficiently, and most grid points will therefore not be constrained by the upper limit.  Sticking to a total column density of $10^{25}\,\text{cm}^{-2}$, there are four grid points which produce observable amounts of CO(40-39) emission: those with $n = 10^{8}$ and $F_{X} = 10^{6}$, $n = 10^{7}$ and $F_{X} = 10^{5}$, $n = 10^{6}$ and $F_{X} = 10^{4}$, and $n = 10^{5}\,\text{cm}^{-3}$ and  $F_{X} = 10^{3}\,\text{erg}\,\text{cm}^{-2}\,\text{s}^{-1}$. We investigate here what sizes the gas structures with these parameters must have to match the CO(40-39) upper limit.  The slab with $n = 10^{5}\,\text{cm}^{-3}$ and $F_{X} = 10^{3}\,\text{erg}\,\text{cm}^{-2}\,\text{s}^{-1}$ is however excluded from 
the analysis, because the CO lines become optically thick, and 
the observed CO emission depends strongly on the position of the observer with respect to the slab. 

Assuming a toroidal gas distribution, with $N_{H} = 10^{25} \text{cm}^{-2}$, we calculate the surface covering factor (1 for a $4\pi\,\text{r}^{2}$ shell covered with gas) that is needed to match the CO(40-39) upper limit. The results are given in Table \ref{tab:gas-mass}, and are interpreted as follows:
\begin{itemize}
 \item A gas layer with $n = 10^{8}\,\text{cm}^{-3}$ at 0.5 pc from the AGN needs a covering factor of 4.5 to be observable, which is not physical. Therefore, such a compact component could not be observed in CO rotational lines at the current sensitivity, even if it has a column high enough to completely block the X-rays from our view.
 \item The slabs placed at 1.6 and 5 pc do give reasonable covering factors of respectively 0.4 and 0.06.
\end{itemize}

MIDI observations of the hot dust in the NGC 1068 nucleus are resolved in two central components with scales of $0.5-3$ pc \citep{Lopez-2014}, and a third component that is offset. If the gas is distributed at scales similar to these central components, the scenario with a radius of 5 pc is not valid. This leaves only one scenario left under which the CO(40-39) could be matched: the one that was presented in Section \ref{sec:slab}.

\begin{deluxetable}{cccccc}
\tablecaption{Matching the CO(40-39) upper limit}

\tablehead{\colhead{Fx} & \colhead{n} & \colhead{Inner radius} & \colhead{Outer radius} & \colhead{Covering factor}  & \colhead{Required emission}\\ 
\colhead{($\text{erg}\,\text{cm}^{-2}\,\text{s}^{-1}$ )} & \colhead{(cm$^{-3}$)} & \colhead{(pc)} & \colhead{(pc)} & \colhead{(0-1)} & \colhead{$\text{erg}\,\text{s}^{-1}\,\text{cm}^{-2}\,\text{sr}^{-1}$} } 

\startdata
$10^{6}$ & $10^{8}$ & 0.5 & 0.53 & 4.5 & 1.18\\
$10^{5}$ & $10^{7}$ & 1.6 & 1.9 & 0.4  & 0.118\\
$10^{4}$ & $10^{6}$ & 5 & 8.2 & 0.06   & 0.0118\\
\enddata

\tablecomments{Surface covering factors required to fit the $\text{J} = 40$ upper limit face value (line absorption by dust is not taken into account). The third column in the table lists the  inner radius of the torus, assuming that $L_{1-100\,\text{keV}} = 10^{43.5}\,\text{erg}\,\text{s}^{-1}$. The outer radius follows from a column of $10^{25}$ cm$^{-2}$.  A $10^{8}$ cm$^{-3}$ density slab located at 0.5 pc from the AGN, needs a covering factor $> 1$ to be observable, which is not physical. The scenarios with $10^{7}$ cm$^{-3}$ and $10^{6}$ cm$^{-3}$  could be observable in CO(40-39).}
\label{tab:gas-mass}
\end{deluxetable}

If the CO(40-39) line were detected at the upper limit value, it would trace gas between 1.6 and 5 pc from the AGN, with densities $\sim 10^{6} - 10^{7}\,\text{cm}^{-3}$, which covers at least 6\% to 40\% of a shell around the AGN. Column densities of $10^{25}\,\text{cm}^{-2}$ are required. Only a few AGN satisfy this condition, and the detection of CO(40-39) in an XDR would point to very high column densities (although this line can be observed at lower column densities in shocks; \citet{Herczeg-2012}). Currently,  there are no instruments which can perform an even deeper observation of this line, but it may be possible with future observatories such as the Space Infrared Telescope for Cosmology and Astrophysics (SPICA).

\section{Summary and Conclusion}
\label{sec:summary}
The deep observation on the CO(40-39) line in the NGC\,1068 central region resulted in an upper limit of $2\times 10^{-17}\, \text{ W m}^{-2}$, more than 10 times lower than the previous upper limit (HD12). The upper limit does not change the previous conclusions, as to what mechanism heats the Highly Excited Component in the CO SLED (HD12): all models (shocks, PDR, XDR) that fit the highest excited component in the observed SLED, are consistent with the new $\text{J} = 40$ upper limit.

We extended an XDR-code, used for protoplanetary disks, with densities up to $10^{9}\,\text{cm}^{-3}$ and input fluxes up to $F_{X} = 10^{6}\,\text{erg}\,\text{cm}^{-2}\,\text{s}^{-1}$ to investigate in how far the upper limit constrains the gas distribution around the AGN. The spectral index of the modeled X-ray spectrum is -1, and the slabs have column densities of $10^{26}\,\text{cm}^{-2}$. In the semi infinite slab model, the X-ray source and the observer are located on the same side of the slab. The observed CO(40-39) flux differs by at most 30\% if the observer is placed at the opposite side of the slab, as long as $N_{H} \leq 10^{25}\,\text{cm}^{-2}$. The difference increases however rapidly for higher column densities. We investigated the CO emission from a gas with density $10^{7}\,\text{cm}^{-3}$, $F_{X} = 10^{5}\,\text{erg}\,\text{cm}^{-2}\,\text{s}^{-1}$, which peaks around $\text{J} = 40$. The modeled CO(40-39) flux matches the upper limit for $N_{H} = 1.5\times 10^{25}\,\text{cm}^{-2}
$.

A line detected at upper limit value would trace gas with densities $\sim 10^{6}-10^{7}\,\text{cm}^{-3}$, at  $1.6-5\,\text{pc}$ from the  AGN, with column densities of at least $10^{25}\,\text{cm}^{-2}$, and surface covering factors of at least $6\%-40\%$. Especially because of the high column density required to be observable within the PACS sensitivity, we expect to see only few AGN with CO(40-39) emission. 

Because tori are thought to be clumpy structures, rather than a smooth medium, we approximated a clumpy distribution with 10 slabs and volume filling factors 1, 0.5, and 0.2. This did not change the CO SLED at CO(40-39), because the increasing layer surface compensates the geometrical dilution. A difference is only seen when $\Phi_{V} = 0.05$. 

On the other hand, line absorption by dust is a very important effect. Dust attenuates the high J CO lines, which makes a difference of a factor 10 in the modeled CO(40-39) line flux at a column density $N_{H} = 10^{25}\,\text{cm}^{-2}$.

In HD12 we speculated that there might be no molecular torus, based on the non-detection of several high-J CO lines. But we did not have an XDR code that addresses all the caveats and uncertainties that are described above. With our new model, adapted  to the high densities and X-ray fluxes in an AGN torus, considering the position of the observer and line absorption by dust, we conclude that the non-detection of CO(40-39) is still consistent with the existence of a molecular torus.

\acknowledgements
We thank Rowin Meijerink for providing us with an extended XDR grid. Basic research in IR astronomy
at NRL is funded by the US ONR; J.F. also acknowledges support from the NHSC. E.G-A is a Research Associate at the Harvard-Smithsonian Center for Astrophysics. A.S. thanks the DFG for support via German-Israeli Project Cooperation grant STE1869/1-1.GE625/15-1. S.V. acknowledges support by NASA through Herschel contracts 1427277 and 1454738.
PACS has been developed by a consortium of institutes led by MPE (Germany) and including UVIE (Austria); KU Leuven, CSL, IMEC (Belgium); CEA, LAM
(France); MPIA (Germany); INAF-IFSI/OAA/OAP/OAT, LENS, SISSA (Italy); IAC (Spain). This development has been supported by the funding agencies
BMVIT (Austria), ESA-PRODEX (Belgium), CEA/CNES (France), DLR (Germany), ASI/INAF (Italy), and CICYT/MCYT (Spain).

\bibliographystyle{apj}

\bibliography{references}

\end{document}